# Automatic Detection and Diagnosis of Biased Online Experiments

Nanyu Chen
LinkedIn Corporation
1000 W Maude Avenue
Sunnyvale, CA, USA
mliu@linkedin.com

Min Liu
LinkedIn Corporation
1000 W Maude Avenue
Sunnyvale, CA, USA
mliu@linkedin.com

Ya Xu
LinkedIn Corporation
1000 W Maude Avenue
Sunnyvale, CA, USA
yaxu@linkedin.com

## ABSTRACT

We have seen a massive growth of online experiments at LinkedIn, and in industry at large. It is now more important than ever to create an intelligent A/B platform that can truly democratize A/B testing by allowing *everyone* to make quality decisions, regardless of their skillset. With the tremendous knowledge base created around experimentation, we are able to mine through historical data, and discover the most common causes for biased experiments. In this paper, we share four of such common causes, and how we build into our A/B testing platform the automatic detection and diagnosis of such root causes. These root causes range from design-imposed bias, self-selection bias, novelty effect and trigger-day effect. We will discuss in detail what each bias is and the scalable algorithm we developed to detect the bias. Surfacing up the existence and root cause of bias automatically for every experiment is an important milestone towards intelligent A/B testing.

## 1   INTRODUCTION

A/B testing, also known as controlled experiment [18,19], has gained popularity as the gold standard for evaluating new product ideas across industry. Many companies have built in-house A/B testing platforms to meet their complex experimentation needs. Some have been discussed at length in past KDD papers [4,5,6,7], including best practices and pitfalls [1,2].

Over the past few years we have seen a massive growth of online experiments at LinkedIn. As a reflection of our 'test everything' mindset, not only has the portfolio of experiments grown to be more diverse, the number of metrics generated for each experiment has more than quadrupled (from 1000 to 4500). While our experimentation platform is powerful and flexible enough to enabled automated design and analysis, it has become apparent that not everyone is able to digest such enormous amount of information and conduct successful A/B tests without guidance. The platform provides 'what' has been impacted but the experimenters are left alone to find out 'why'.

The goal of our work is to fill such gap by *automatically* surfacing up important information experimenters should know and guiding them towards answers they look for. This enables a scalable path towards further growth around experimentation. Moreover, many of these insights are discovered through mining a large number of historical experiments collectively. Such meta-analysis can be more powerful and insightful than individual experiment analysis in isolation.

One of the biggest applications of insights is to uncover why an experiment is biased. The bias problem is particularly prevalent for triggered analysis [9], which limits the analysis to only affected users and is widely accepted to be more sensitive. At LinkedIn, about 10% of our triggered experiments used to suffer from bias. Needless to say, it is a nightmare to get biased results as it renders the whole experiment in vain. It is even worse when we do not know how to fix it. After analyzing a large number of biased experiments, we are able to summarize and categorize the insights into an automated toolkit, which significantly reduced investigation time from weeks to within hours.

In the analysis stage, the challenging 'why' component comes in two aspects. First, when metrics are impacted unexpectedly, one often needs to find out 'why' and attempts to do so by going through the 4000+ metrics computed to find clues. Second, one usually expects consistent impact from experiments. The reality is that there are many occasions the size of the impact will evolve in time. One either observes such scenario or gets puzzled by 'why' the impact is changing or entirely misses such change, especially when it happens in metrics that are not actively sought after. In section 3.2 we introduce meta-analyzing the correlation of their impact in experiments to identify metrics that are likely to move together, through which we can help to form hypothesis on 'why' metrics are impacted. In Sections 3.3 and 3.4 we define, formulate and introduce detection algorithms on two causes of time-dependent impact, named as trigger-day effect and novelty effect [10]. Even though some past papers [1,2] have given practical examples of such effect, they were based on heuristics and none discuss how to automatically detect such effect. Not only was it a non-trivial problem to identify and formulate these different causes of time related changes, one needs to deal with statistical noise and be able to identify true signal as well as validate it. When we productionized these algorithms, we had to leverage existing summary statistics as much as possible so that we can serve these insights timely without adding huge costs to the system.

Here is a summary of our contributions in this paper:

- As far as we know, we are the first to productionize automated algorithms to mine insights using meta-analysis on historical experiment data.
- The four insights we share are generally applicable. The methodologies can be applied to meta-analyzing historical experiments from any A/B testing platform, and the production of such methodologies into automatic insights discovery engines is inexpensive as it uses summary statistics that are computed by most A/B testing platform.
- We study extensively the benefit of triggered analysis, and for the first time, we show that some experiments can appear

- to have burn-in effect purely because of the triggering mechanism.
- We share many real experiments from LinkedIn where design tends to be biased and present learnings from debugging such biased designs.

## 2 Triggered Analysis

Some of the challenges we present arise particularly in triggered analysis, where only users who were actually impacted by the experiment are included to help separate signal from noise [11]. For this reason we devote this section to discuss it in depth.

Instead of session-triggered analysis [11], we focus on user-triggered analysis which is more widely used as it does not assume that treatment effect only exists in the sessions that users trigger. Triggered users are usually identified when experiment assignment evaluation code is called [7]. Once a user is triggered in an experiment, all activities in a given analysis period are counted, even on days users do not trigger. This leads to an interesting classification of metrics that turned out to have unique properties in experimental analysis. Each metric belongs to one of the following two categories relative to the trigger condition: 1) *fully-covered metrics*: metrics that are entirely nested under the trigger condition, and 2) *partially-covered metrics*: metrics that are only partially covered by the trigger condition. In an experiment triggering on landing the profile page, profile views is a fully-covered metric as there is no way one can have a profile view without landing on the profile page. Non-triggered users all possess zero values in such metrics and triggered users also possess zero values on days they do not trigger. On the other hand, total page view is a partially-covered metric as one can contribute values to this metric through profile views as well as other channels such as home page. As a result, users can still have engagement on days they do not trigger. The fact that experiments can have different impact in these metrics on days users trigger (in-trigger impact) and otherwise (off-trigger impact) has lead to a profound observation we made, named as trigger-day effect, and will be discussed thoroughly in section 3.3.

While it is a common intuition that triggered analysis helps bring up signal from noise, no past work has extensively discussed the details. We show why triggered analysis is beneficial by evaluating the consequences of non-triggered analysis, where users that are not affected by the experiment are included (such as all-user analysis).

We assume the following notation:

|  | Triggered Treatment | Triggered Control | All User Treatment | All User Control |
|---|---|---|---|---|
| Metric Sum | $\Sigma_t$ | $\Sigma_c$ | $\Sigma_t'$ | $\Sigma_c'$ |
| Sample Size | $n_t$ | $n_c$ | $n_t'$ | $n_c'$ |
| Variance | $var_t$ | $var_c$ | $var_t'$ | $var_c'$ |

Also, let $k = n_c'/n_c$. $r = n_t/n_c = n_t'/n_c'$. For business interpretation purpose we usually test and report if the percentage lift $\Delta\% = \overline{X_t}/\overline{X_c} - 1$ is 0 based on the test statistic $t = \frac{\Delta\%}{\sqrt{var(\Delta\%)}}$, where $\overline{X_t} = \frac{\Sigma_t}{n_t}$ and similar definition applies to $\overline{X_c}$. One can show that $var(\Delta\%) = \frac{var_t}{\overline{X_c}^2 n_t} + \frac{var_c \overline{X_t}^2}{\overline{X_c}^4 n_c}$ based on Delta method [13].

*Fully-covered Metrics*: As aforementioned, users who would not have triggered will not have non-zero values. Hence

$$E(\Delta\%') = E\left(\frac{\Sigma_t'/n_t'}{\Sigma_c'/n_c'} - 1\right) = E\left(\frac{\Sigma_t/n_t}{\Sigma_c/n_c} - 1\right) = E(\Delta\%)$$

And

$$\frac{var(\Delta\%')}{var(\Delta\%)} = \left(\frac{var_t' + r(1+\Delta\%)^2 var_c'}{var_t + r(1+\Delta\%)^2 var_c}\right)k$$

Note that $var_t' = \frac{1}{k}\left[var_t + \left(1 - \frac{1}{k}\right)\overline{X_t}^2\right]$ (similarly for $var_c'$). Hence

$$\frac{var(\Delta\%')}{var(\Delta\%)} = 1 + \frac{\overline{X_T}^2 + r(1+\Delta\%)^2\overline{X_C}^2}{var_t + r(1+\Delta\%)^2 var_c}\left(1 - \frac{1}{k}\right) > 1$$

One can see that variance is always bigger in non-triggered analysis. The higher the $k$, the smaller the coefficient of variation of the metric, the more inflated the variance (and hence the less powerful the analysis).

*Partially-covered Metrics:* We can no longer assume $\Sigma_c' = \Sigma_c$. Instead, let $s = \Sigma_c'/\Sigma_c$ which is greater than 1 by definition. For simplicity, we also assume $var_t = var_c, var_t' = var_c'$, and Let $r_\sigma = var_c'/var_c$ be the variance inflation ratio. Practically, triggered users tend to be more homogeneous than the general population because they all satisfy the trigger condition, hence variance of triggered users tend to be smaller than that of the population, i.e. $r_\sigma \geq 1$. Unlike fully-covered metrics, the effect size on the overall population is diluted. In fact, $E(\Delta\%') = E(\Delta\%)/s$. Note also that $\overline{X_c'}/\overline{X_c} = s/k$. One can see by plugging in the variance formula that

$$\frac{var(\Delta\%')}{var(\Delta\%)} = \left(\frac{1 + r(1+\Delta\%/s)^2}{1 + r(1+\Delta\%)^2}\right)\frac{kr_\sigma}{s^2}$$

Hence the ratio between the population and triggered t statistics is,

$$\frac{t'}{t} = \left(\frac{1 + r(1+\Delta\%)^2}{1 + r\left(1 + \frac{\Delta\%}{s}\right)^2}\right)^{1/2} \frac{1}{kr_\sigma^{1/2}}$$
$$< \frac{1 + \Delta\%}{1 + \frac{\Delta\%}{s}} \frac{1}{kr_\sigma^{\frac{1}{2}}}$$
$$\approx \frac{1}{kr_\sigma^{\frac{1}{2}}}$$

The inequality holds because $\frac{t'}{t}$ is monotone in $r$, the approximation holds because $\Delta\%$ is usually a very small number, e.g. $< 5\%$, therefore $\frac{1+\Delta\%}{1+\frac{\Delta\%}{s}} \approx 1$. On the other hand,

$$kr_\sigma^{\frac{1}{2}} = \frac{n_c'}{n_c}\sqrt{\frac{var_c'}{var_c}}$$
$$= \sqrt{\frac{n_c'^2 var_c'}{n_c^2 var_c}}$$
$$= \sqrt{\frac{n_c' SS_c'}{n_c SS_c}}$$

where $SS_c$ is the sum of squares of the metric for triggered members, and $SS_c'$ is that for the population. Obviously $SS_c' > SS_c$ and $n_c' > n_c$, so $kr_\sigma^{\frac{1}{2}} > 1$. Therefore, $\frac{t'}{t} < 1$, that is the t statistic in non-triggered analysis is always smaller, hence weaker signal.

## 3 Insightful Discoveries

We now share the four insightful discoveries we learnt through uplifting our experimentation platform.

We use superscripts to denote analysis date range. $n^{[x,y]}$, $\Delta\%^{[x,y]}$ represent the sample size and percentage impact of an experiment between day x and day y inclusive, respectively. On day k of an active experiment, the analysis pipeline computes summary statistics for cross-day [1,k-1] that aggregates all data throughout the experiment and latest single-day [k-1, k-1] by default and other date range combinations are only available on request.

## 3.1 Diagnosing Biased Experiments

One of the assumptions of Rubin's causal model [12] is Ignorable Treatment Assignment Assumption (also known as *unconfoundedness*). In particular, a user's inclusion in an experiment should be independent from the treatment assignment. This is rarely a challenge for experimenters who can pre-fix assignment to experiment targets that are fully randomized. However, in triggered analysis, samples are collected sequentially as users conduct activities on the site and the unconfoundedness assumption become harder to control. One can efficiently identify if there is bias in the triggered sample by running chi-squared test of goodness of fit. Suppose we observe sample sizes $n_t$ and $n_c$ in treatment and control in an experiment with traffic allocation $r_t$ and $r_c$. For simplicity we assume there are only two variants in the experiment hence $r_c = 1 - r_t$. Also let $E_t = (n_t + n_c)r_t$ and $E_c = (n_t + n_c)r_c$ be the expected sample sizes. One can test whether the observed distribution matches expected distribution using the chi-squared statistic $\frac{(n_t - E_t)^2}{E_t} + \frac{(n_c - E_c)^2}{E_c}$, which has a $\chi_1^2$ distribution under the null hypothesis. We will refer to this test as 'Sample Size Ratio Test' from now on.

The real challenge, however, turned out to be knowing what causes the bias when the Sample Size Ratio Test is rejected. In this section we share our lessons learnt through building an automated toolkit to identify root causes, which significantly reduced diagnosis time from weeks to within hours. To identify what components needed to go into the toolkit, we had to first investigate the root cause of a large number of experiments one at a time as well as classifying the unique causes and their characteristics. Some of these challenges are specific to triggered analysis but it is still the preferred analysis whenever possible. One may refer back to section 2 for a fully-fledged discussion.

### 3.1.1 Examples

*Dynamic Targeting:* Targeting [7] refers to running experiments on specific user sets based on their properties and activities to deliver personalized experience. While most targeting criteria are static for our users, such as country, industry and locale, we also use more dynamic targeting criteria that can change on a regular basis. One needs to consider whether a treatment itself would interact with the targeting criteria and hence would result in users switching in/out of an experiment segment at different rates in different variants. To illustrate this with an example, one of the widest applications of recommendation systems at LinkedIn is Jobs You May Be Interested In (JYMBII) [16]. Different algorithms are used to recommend jobs based on member attributes such as whether they are students or executive as well as their location. In particular, we have learnt that it is necessary to have different algorithms for active job seekers and passive job seekers. On the other hand, a separate machine-learning algorithm is used to classify whether members are actively seeking jobs. We ran experiments on passive job seekers and over time a better performing model had proportionally less users triggered. It turned out that the model lifted page views and clicks on jobs pages, and such metrics, as part of the classification model, removed some members from the targeted experiment. Such feedback loop is likely to cause bias in estimation of treatment effect. We have seen many similar experiments of this kind, such as experiments urging members to complete profile running on members with incomplete profile and experiments re-engaging dormant members.

*Cool-off:* Among the many features we A/B test on, there is a set of cases where we would like to cap the number of impressions a member has on certain pages or widgets within a period of time ("cool-off" period). The most common situation like this is called "cross promotion" [14], a common marketing strategy for companies to target customers of a product for a related product or functionality. For online companies like LinkedIn, these are usually widgets embedded within certain parts of websites and native apps that recommend members to engage in related features. However, such widgets always result in two-sided effects: while the promoted product is expected to gain engagement, members are likely distracted from the page the cross promotion is placed on. As a concrete example, LinkedIn recently launched a feature to allow members to rebuild their own feed by following content they care about. We ran an A/B test on whether or not to put a cross promotion of such feature at the top of the feed page to measure its effectiveness as well as its implication on feed engagement metrics. Initially, this cross promotion experiment was implemented with the logic in figure 1 left. In this case, the cool off condition is (impression > 2 or click > 0). The seemingly correct logic resulted in severe sample size ratio mismatch. As described in section 2, triggered analysis relies on runtime tracking events that are fired during variant evaluation (line (3)). The logic above does not fire tracking events when a member is cooled off. Since only members in treatment group will ever be in cool-off period after conditions are met, there will be proportionally less unique users in treatment group unless we count members from the very beginning of the experiment. As with any other experiments, it is best practice to go through a few iterations starting from a small traffic percentage to mitigate risk. It is hence impractical to be able to always count traffic from the very beginning. Note that the analysis based on mismatched users from logic above are usually negatively biased since users who are more active are more likely to trigger into experiments early on, and hence are more likely to be excluded from the analysis report. The solution to this problem is to exchange lines (1) and (3).

```
If (member in cool-off):        (1)
    Do not show widget          (2)
Else if (member in control):    (3)
    Do not show widget          (4)
Else:                           (5)
    Show widget                 (6)
```

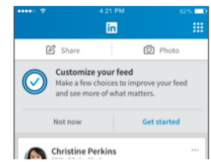

Figure 1: Left: cool-off biased code. Right: cross promotion example

*Residual Effect:* The residual effect usually refers to contamination of a former experiment to a subsequent one with the same user split. This can happen between different ramps of the same experiment, or within the same experiment if we start counting samples from a time point later than the activation of the experiment. When the treatment effect is large, it may change the frequency at which users visit. As a result, residual effect may lead to mismatched sample sizes. One such example was with our People You May Know relevance algorithm improvement. The experiment showed promising results on engagement across the site but started showing sample size ratio mismatch after the first ramp. It turned out that this was purely due to the fact that the algorithm was so good that it made users come back more often! Such bias can be corrected by re-randomization and counting users from the very beginning of the experiment. However, from our experience it is rare that treatment itself would be impactful enough to change user re-trigger rate.

*Biased implementation* can create similar symptoms as if residual effect is caused by treatment itself. As an example, when we re-vamped LinkedIn homepage, the treatment evaluation code was written in two services: 1) When users hit the router by typing 'linkedin.com'. 2) When users enter the new homepage directly with designated Urls. The second code call was needed so that if the experiment was terminated, users cannot access the new homepage directly with the Urls. It turned out that we saw more than expected users in treatment after the first ramp whereas including users from the first experiment will resolve the bias. It turned out that this was merely due to the fact that some users enter the site through direct Urls (without going through the router) and only those in treatment group would be evaluated in such case, resulting in additional user count.

*Dependent Experiments*: By using the same hash ID, one effectively creates a layer in which dependent tests can be tested in fully or fractionally factorial fashion [7]. However, when such dependent tests have chronological dependency (instead of all at the same time), the child page experiment's trigger rate can change as a result of treatment effect of experiments earlier on. As an example, we recently tested some new page designs on our job posting page (the parent page) and the checkout page (the child page). The main change we made on the posting page is hiding the price and the checkout page was redesigned with more clarity and simplicity. The team implemented this change with a separate experiment on the posting page and on the checkout page triggering on landing the corresponding pages and avoided the hide price/old design combination, which will result in ambiguity in pricing. The checkout page experiment showed mismatched sample sizes with more than expected member count in the treatment group and the key success metrics such as job booking was shown as directionally different from what was suggested based on the job posting page experiment. It turned out that this was purely because the parent page experiment changed the CTR to the child page. In more complex situations such chronological dependency in trigger condition can happen mutually between experiments and result in biases in both.

*2.1.2 Generalization for Diagnosis*

The examples shared in the previous section were categorized through our effort to diagnose many experiments manually one at a time. With all our learning we have since then built a toolkit that expedited the diagnosis process significantly. When Sample Size Ratio Test fails, we run the following checks automatically:

- Run Sample Size Ratio Test on all targeted users. The reason behind this is obvious so that one can always identify dynamic targeting has resulted in bias.
- Compute sample size ratios separately for users who trigger for the first-time and returned users. This will allow us to separate bias is in the feedback loop, such as cool-off, residual effect and biased implementation from explicit biases in code, such as dependent experiments. New users being unbiased also suggests that rehashing and counting users from the beginning will result in matched samples. Note that one does not need to do explicit aggregation to compute new user counts and returned user count. As mentioned earlier, on day $k$ of an experiment, we have $n^{[1,k-1]}, n^{[1,k]}, n^{[k,k]}$ by default and one can simply conclude $n_{returned} = n^{[1,k-1]} + n^{[k,k]} - kn^{[1,k]}$ and $n_{new} = n^{[1,k]} - 1n^{[1,k-1]}$. Figure 2 shows the new/returned sample size ratios of the People You May Know experiment described in the previous section. As one can see, returned user is statistically significantly mismatched from the second day and provided faster signal .

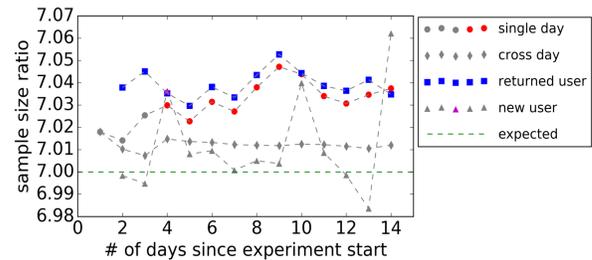

**Figure 1: Sample Size Ratio for New and Returned Users**

- Identify tracking events that are independent of triggering code calls that would reproduce the desired trigger condition. For example, one can utilize page view tracking to identify users landing on particular pages. Such test would allow us to separate whether the bias comes from experiment implementation, such as the cool-off and the biased implementation cases, or is actually a reflection of user engagement biases, such as biased implementation and dependent experiments.
- Track Metadata to separate code calls at different places. In our case, code is divided up into modules that generally map to service names. Splitting triggered traffic by service allows us to identify which part of the code to look into and can usually help us identify biased implementation.
- For each experiment sharing hashID (experiment 1) with the problematic experiment (experiment 2), identify users that trigger both experiments (Set A) as well as in each one only (Set B1 and B2). Ideally, one also needs to split A into users that trigger each of the two experiments first (A1, A2). While experiment 1 has triggered population B1+A1+A2, the only component that can be affected by chronological dependency with experiment 2 is A2. One can perform Sample Size Ratio Test on this set to identify biased dependencies. Practically, splitting A1 and A2 is a challenging process and we have found it useful to start with checking if the overall de-duplicated population A1+A2+B1+B2 is mismatched.

## 3.2 Identifying Related Metrics

For companies running experiments at scale, historical data consisting of thousands of experimental results serve as a valuable asset to learn about metric properties. Deng [17] talked about using historical A/B test data to objectively learn priors on how likely metrics move. In this section we discuss algorithms to learn related metrics that are likely to have correlated treatment effect.

There are two motivations of such analysis. First, it helps with understanding why metrics move. Note that we are not trying to construct formal causal relationships between metrics but rather through correlational analysis forming hypothesis on possible relationships. Second, by leveraging these relationships we may be able to identify 'early indicators' of metrics lacking good statistical properties. In other words, we can form earlier belief on potential impact of a metric based on the experiment's impact on another one. We discuss both applications in this section.

*2.2.1 Meta-analysis: identifying related metrics and how they move together.* To make conclusion based on large-scale historical experimental data one of the fundamental problems one needs to address is noise. In order to identify metrics that move together, a simple method that utilizes the outcomes of hypothesis testing is to count the proportion of times a metric is statistically significant when the other metric is also statistically significantly moved. However, there are both precision and recall problems to such naive counting method:

*Precision:* The Type I error rate is always 5% by definition. This problem amplifies when the metrics (or measures) in question are correlated on the user level. To illustrate the problem, assume $(X,Y) \sim MVN(\mathbf{0}, \Sigma)$ where $\Sigma = \begin{pmatrix} \sigma_x & \rho\sigma_x\sigma_y \\ \rho\sigma_x\sigma_y & \sigma_y \end{pmatrix}$ and $\rho$ is the Pearson's correlation of the two metrics. Figure 3 Right plots $p = P(Reject\ H_0^X\ |\ Reject\ H_0^Y)$ with respect to $\rho$ based on simulation. One can see that the proportion of false positive grows in $\rho$.

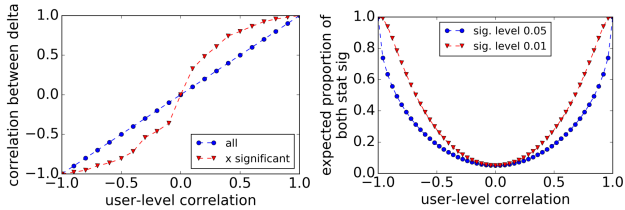

**Figure 3: Left: Spurious Correlation between delta. Right: Expected proportion of both statistically significant**

*Recall:* As in the standard hypothesis-testing framework, Type II error can be optimized but not eliminated. When it comes to learning the relationships of two metrics, the relative Minimum Detectable Effect (MDE, the smallest effect achieving 80% power) of the two metrics may be different from the actual relative treatment effect. In such case even if all experiments impacted both metrics, we will end up with $p < 1$. To make it more complex, relative MDE of real business metrics also depends on experiment length. While most metrics do not have much variance reduction beyond the first week, some benefit from considerable variance reduction over a longer period. Looking at longer running experiments usually improve the recall in identifying relationships with such metrics.

Besides whether the two metrics are likely to move together, one may also be interested in learning relationships in direction and magnitude of treatment effect. One can simply compute the correlation between the effect sizes of the two metrics. However, without false discovery control, one will include data points with spurious correlation as shown in Figure 3 Left.

To mitigate the precision and recall problems described above, we propose the following steps to identify whether and how metrics Y moves when metric X moves:

1. Select a cohort of historical experiments that represent good institutional knowledge. For each experiment use data from the longest available cross-day results in the iteration that has the maximum power and run sufficiently long (for example, at least a week). This is to ensure that the data points considered are independent.
2. Compute the user level correlation $\rho$ between the two metrics and the resulting expected proportion of times the two metrics are both statistically significant under null space (as shown in Figure 3 right). Use chi-squared test to identify if the observed proportion is higher than expected. To identify the metrics that are most likely to move with a particular metric, we can use the chi-squared statistic as a ranking criterion.
3. Run Benjamini-Hochberg algorithm [15] to identify a set of experiments that impacted metric X with false discovery rate controlled. This is to ensure the proportion of data points coming from the null space is small to avoid spurious correlation.
4. Run linear regression of $\Delta_Y = \beta_0 + \beta_1 \Delta_X$ based on the set of experiments identified in step 3. Use $\widehat{\beta_1}$ to estimate how much we expect to move metric Y when we move metric X if the linear relationship is strong. One may need to run outlier removal procedures to ensure the correlation learnt are robust.

*2.2.2 Application: Early Indicator.* When a metric is not statistically significant but underpowered, especially in the earlier stage of the experiment, one often has no clue weather such metric is actually impacted or not. However, leveraging the experiment's impact on metrics that are likely to move with such metric, one can be better informed on the potential impact. In this section we describe a Bayesian algorithm that induces the likelihood metric Y is impacted by an experiment given data on metric Y itself and its 'early indicator' metric X, a metric that is likely to move with Y and has good statistical properties. We compute $P(H_1^Y|\delta_X, \delta_Y)$ using Bayes Theorem

$$\frac{P(H_1^Y|\delta_X, \delta_Y)}{P(H_0^Y|\delta_X, \delta_Y)} = \frac{p(\delta_Y|H_1^Y, \delta_X)}{p(\delta_Y|H_0^Y, \delta_X)} \frac{P(H_1^Y|\delta_X)}{P(H_0^Y|\delta_X)}$$

where

- $P(H^Y|\delta_X) = P(H^Y|H_1^X)P(H_1^X|\delta_X) + P(H^Y|H_0^X)P(H_0^X|\delta_X)$ where $P(H^X|H^Y)$ can be approximated by the proportions learnt from step 2 in the previous section. $P(H_1^Y), P(H_1^X)$ and $P(H_1^Y|\delta_Y)$ are learnt using the Objective Bayesian method described in [17]. $\delta = \Delta/\sigma$ is the normalized impact and $\sigma^2/N_e = \sigma_t^2/N_t + \sigma_c^2/N_c$ and $N_e = 1/(1/N_t + 1/N_c)$.
- $p(\delta_Y|H^Y, \delta_X) = p(\delta_Y|H^Y)$ is simply the likelihood function under the corresponding hypothesis. As in the Objective Bayesian framework, we assume two group model so $p(\delta_X|H_0^X) = \phi(\delta_X, 0, N_e^{pred})$ and $p(\delta_X|H_1^X) = \phi(\delta_X, 0, N_e^{pred} + V_X^2)$ where $\phi(x, \mu, \sigma^2)$ is the normal density function with mean $\mu$ and variance $\sigma^2$. $V_X^2$ is learnt from historical experiments with EM Algorithm. Note that $N_e^{pred}$ is the predicted effective sample size when the experiment runs sufficiently long (e.g. 30 days).

We flag metric impact on metric Y if $P(H_1^Y|\delta_X, \delta_Y)$ is sufficiently large (e.g. $P(H_1^X|\delta_X, \delta_Y) > 0.6$). We evaluated this

algorithm by applying it to day seven results of experiments that ran sufficiently long (at least three weeks). A flagged experiment is considered as 'true positive' if metric Y becomes statistically significant eventually and 'false positive' otherwise. The overall precision was 0.9 and recall was 0.6.

### 3.3 Trigger-day Effect

Starting from this section we discuss causes and detection algorithms for time-dependent treatment effect. Making quality decisions hinges on accurate estimate of treatment impact. When the impact itself is time dependent, it is crucial that experimenters are alerted so that they do not conclude on treatment impact before it stabilizes. Furthermore, experimenters need to know why impact is changing over time as it often reveals valuable insights on how users interact with the treatment. We start with trigger-day effect, an intriguing phenomenon that has caused confusion amongst even the finest data scientists, and then go on to discuss novelty/delayed effect.

As mentioned earlier, for every experiment both cross-day impact and single-day impact are computed. These two impacts usually align in magnitude (see Figure 4 left for an example). However, there exist scenarios where they are drastically different (Figure 4 right). Under such scenario, cross-day impact would also exhibit a strong and consistent trend as experiment progresses. This phenomenon is so puzzling that it often prompts experimenters to suspect a bug in the experimentation platform and question what really is the experiment impact.

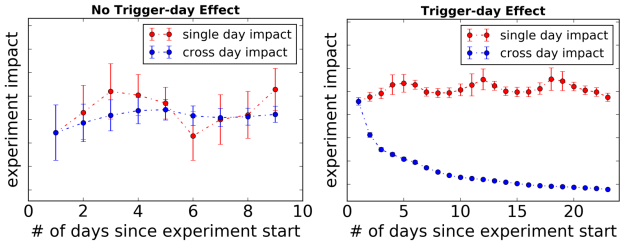

**Figure 4: Left: Example experiment with no trigger-day effect. Right: Example experiment with trigger-day effect.**

We named such phenomenon 'trigger-day effect' because it arises from the nature of triggered analysis. In user-trigger analysis, single-day impact roughly equals in-trigger impact (defined in Section 2) as the single-day analysis population is just the users who triggered on that day. On the other hand, cross-day impact captures all activities of triggered users in the analysis period, including those occurred on days the users do not trigger. It is hence natural to de-compose the cross-day impact into in-trigger and off-trigger impact. Intuitively, in-trigger impact and off-trigger impact can be different. For example, pushing a connection recommendation to users may increase number of connection invitation sent by 100% on the day of the push (100% in-trigger impact), but might have close to zero impact on days a user doesn't receive such push (0% off-trigger impact). As a result, single-day impact might be different from cross-day impact as the former equals in-trigger impact and the latter is a mix between in-trigger and off-trigger impact. And the time dependency of cross-day impact results from changing composition of the two types of impact.

To understand trigger-day effect beyond heuristics, we now formulate it mathematically. Let $X_i$ denote the total metric value of user $i$ in the experiment, $I_i$ be the metric value from triggered days and $O_i$ from non-triggered days. Hence $X_i = I_i + O_i$. For an experiment running for $k$ days,

$$\Delta\%_X^{[1,k]} = \overline{X_t}^{[1,k]} / \overline{X_c}^{[1,k]} - 1 = w_k \Delta\%_I + (1 - w_k)\Delta\%_O$$

where $\Delta\%_I^{[1,k]} = \overline{I_t}^{[1,k]} / \overline{I_c}^{[1,k]} - 1$ is the *in-trigger* impact, $\Delta\%_O^{[1,k]} = \overline{O_t}^{[1,k]} / \overline{O_c}^{[1,k]} - 1$ is the *off-trigger* impact and $w_k = \overline{I_c}^{[1,k]} / \overline{X_c}^{[1,k]}$ is the fraction of in-trigger contribution ($0 \leq w_k \leq 1$).

A couple of important observations:

1. The in-trigger impact, $\Delta\%_I^{[1,k]}$, is time independent and is by expectation equal to single-day impact $\Delta\%_X^{[t,t]}$ for $t = 1, \ldots k$ since single-day analysis population is by definition users who triggered on that day. Similarly, $\Delta\%_O^{[1,k]}$ is also time independent, and denoted by $\Delta\%_O$.

2. Cross-day impact $\Delta\%_X^{[1,k]}$ is a weighted average of in-trigger and off-trigger impact. It is expected to differ from in-trigger impact, hence single-day impact, if 1) in-trigger impact and off-trigger impact are different and 2) off-trigger impact has a non-zero share in the cross-day impact, i.e. $w < 1$. This is equivalent of metric being partially-covered, as defined in Section 2.

3. Cross-day impact $\Delta\%_X^{[1,k]}$ is time dependent because $w_k$ is. Suppose on any given day a user triggers with probability p. Let $r = I/O$ where $I$ and $N$ are expected contributions to metric value from a trigger day and a non-trigger day. For an experiment that ran for $k$ days and a total of $n$ users have triggered, we expect $n\binom{k}{t}p^t(1-p)^{k-t}$ users to have triggered on $t$ days. Therefore,

$$\Sigma_i I_i = \sum_t n\binom{k}{t}p^t(1-p)^{k-t} tI = npkI$$

$$\Sigma_i O_i = \sum_t n\binom{k}{t}p^t(1-p)^{k-t}(k-t)O$$

$$= np[(1-p) - (1-p)^k]O$$

Hence $w_k = \frac{\Sigma I_i}{\Sigma I_i + \Sigma O_i} = \frac{pr}{(1-p)-(1-p)^k+pr}$. This model captures well how $w$ evolves in an experiment. Figure 5 compares the real change in $w$ with the theoretical change in $w$ modeled by above expression using data from the experiment shown in Figure 4 right.

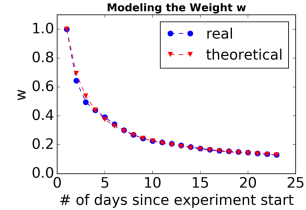

**Figure 5: w evolving over time**

Note that not only does this model explain why $w$ is time-dependent, it also projects what the cross-day impact would stabilize to. As $k \to \infty$, $w \to \frac{pr}{1-p+pr}$, which means cross-day impact will be only $\frac{pr}{1-p+pr}$ of the initial impact after it stabilizes.

With observations above we designed and algorithm to automatically flag trigger-day effect when the following conditions are met:

1. $w$ is reasonably small. for reasons discussed in point 2 of 'important observations'.
[1] $\Delta\%_I^{[1,k]}$ and $\Delta\%_O^{[1,k]}$ are statistically significantly different. Note that $\Delta\%_I^{[1,k]}$ and $\Delta\%_O^{[1,k]}$ can be derived from existing cross-day and single-day summary statistics, but not $var_I$ or $var_O$. However, note that if $cov(I, O) \geq 0$, then $var(X) = var(I + O) = var_I + var_O + 2cov(I, O) \geq var_I$ (and $var_O$). Therefore we can bound $var(\Delta\%_T^{[1,k]})$ and $var(\Delta\%_N^{[1,k]})$, and use such variance estimate to perform a t-test between $\Delta\%_I^{[1,k]}$ and $\Delta\%_O^{[1,k]}$.

## 3.4 Novelty Effect

Another cause for time dependent treatment impact is novelty effect. Unlike trigger-day effect, novelty effect does not stem from the nature of triggered analysis, but rather from user behavior change. Essentially, a user's reaction to a treatment can be different the first time she triggers it, vs. when she has triggered it many times. For example, users might be very responsive to notifications at first, but over time they learn to ignore or disable them, thus making the treatment less effective. When novelty effect exists, we need to automatically surface such information to experimenters because not only does it indicate the experiment needs to run longer to get a stable impact estimate, it also offers unique insights on how users interact with the new feature being experimented on.

The fact that the treatment gets less/more effective as users trigger it more means we should expect to see a strong and consistent trend in single-day impact when the treatment experiences novelty effect (see Figure 6 left for an example). Reason being, if we take the single-day analysis population and compute the average number of days the experiment was triggered, we should expect that number to grow as experiment progresses (see Figure 6 right for the average number of days triggered computed from the experiment in Figure 6 left).

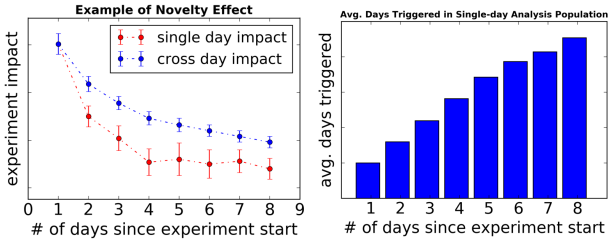

**Figure 6: Example of Novelty Effect**

With the observations above, we propose the following algorithm to detect novelty effect:

1. Run linear regression $\Delta\%^{[t,t]} = \beta_0 + \beta_1 \frac{1}{t^\alpha} + \beta_2 \frac{1}{t^\gamma}$ with one week's single-day impacts. $\alpha$ and $\gamma$ are chosen so that $\frac{1}{t^\alpha}$ is a slowly-decaying term with respect to t while $\frac{1}{t^\gamma}$ is a fast-decaying term (for example, $\alpha = 0.35, \gamma = 2$). These two terms were chosen to represent the two typical types of trend we identified, as illustrated in figure 7, through analyzing the single-day impact trend in thousands of real experiments. The slowly decaying term fits the gradual type of trend on the left, while the fast decaying term fits well with the 'elbow' type on the right. A linear combination of the two captures trends in between.

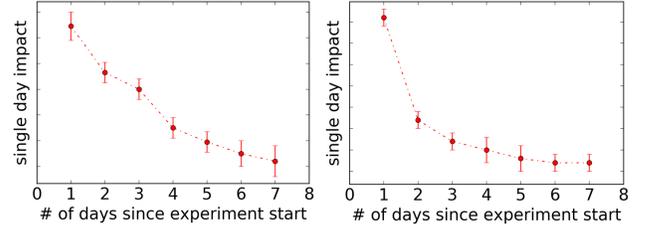

**Figure 7: Two Typical Types of Single-day Impact Trend**

2. We flag novelty effect when all the following conditions are met: 1) The linear model captures the single-day effect trend well. (for example, $R^2 \geq 0.8$). 2) The fitted line is monotonic in $t$. 3) The largest single-day impact is statistically significantly different from smallest single-day impact. A smaller significance level (e.g. at 0.005) is helpful in limiting the number of false positives. This condition has proven to be effective in isolating true novelty effect from spurious ones [1].

We have identified many experiments with novelty effect using this algorithm. In most cases the experimenters completely did not their experiment would have novelty effect. Based on all the novelty effects identified, we have noticed three types of experiments that are most prone to novelty effect: 1) Big changes where users tend to explore the new features in the initial days. 2) Notification or badging experiments 3) Recommendation algorithms with limited candidate pool, for example, Connections Updates and People You May Know.

Finally, we would like to point out a caveat of this algorithm and some additional analysis to complement it:

- The algorithm does not distinguish novelty effect from other time-dependent effects that may share the same trend. In particular, we have seen examples of day-of-week effect where weekend impact differs from weekday impact. When such experiments are activated on weekends, the single-day trend resembles novelty effect very much.
- It is impossible to estimate the magnitude of novelty effect based on a sequence of single-day impacts alone, because on any given day, the experiment population is a mix of users triggering the experiment for the first time, as well as users who have triggered it many times. When measuring novelty effect itself is important, one can divide the treatment population into two subpopulations. Expose one of them to the treatment first, wait till treatment impact is stable. Then expose the second subpopulation to treatment. The difference in treatment impact between the subpopulation experiencing treatment for the first time vs. the subpopulation that has experienced the treatment many times is just the magnitude of novelty effect.

## 4 Summary and Future Work

In this paper, we shared four insightful discoveries we learnt through meta-analyzing historical A/B tests. We presented common ways in which experiments are biased and proposed a diagnosis framework. To help explain why metrics move, we also presented our approach to leverage experiment meta-analysis to identify related metrics, and leverage such method to identify early indicators of key metrics lacking good statistical property. We also shared a few algorithms to detect effect changing over time.

An area we have not discussed about in this work is identifying heterogeneous treatment effect. Such analysis is particularly challenging in multiple and high dimensions. Finally, we hope our work will encourage more research in mining experimental data to build smarter A/B testing platform that will eventually result in "Artificial Intelligence" in decision-making.

## ACKNOWLEDGMENTS

The authors would like to thank XXX for insightful discussions on examples shared in this paper. We would also like to thank all members of the experimentation team who have made it possible to build many of the insights discussed in this work in production.